\documentclass[10pt,twocolumn,a4paper]{article}

\usepackage[utf8]{inputenc}
\usepackage[T1]{fontenc}
\usepackage{amsmath,amssymb,amsfonts,amsthm}
\usepackage{graphicx}
\usepackage{booktabs}
\usepackage{multirow}
\usepackage{algorithm}
\usepackage{algorithmic}
\usepackage{xcolor}
\usepackage[margin=0.75in,columnsep=0.25in]{geometry}
\usepackage{float}
\usepackage{caption}
\usepackage{subcaption}
\usepackage[numbers,sort&compress,square]{natbib}
\usepackage{setspace}
\usepackage{fancyhdr}
\usepackage{microtype}
\usepackage{titlesec}
\usepackage{newtxtext,newtxmath}

\titleformat{\section}
  {\normalfont\large\bfseries}{\thesection}{1em}{}
\titleformat{\subsection}
  {\normalfont\normalsize\bfseries}{\thesubsection}{1em}{}

\usepackage[colorlinks=true, linkcolor=blue, citecolor=blue, urlcolor=blue]{hyperref}

\newtheorem{proposition}{Proposition}

\newtheorem{definition}{Definition}
\newtheorem{corollary}{Corollary}
\newtheorem{remark}{Remark}

\title{\textbf{Covariance-Aware Simplex Projection for Cardinality-Constrained Portfolio Optimization}}

\author{
  Nikolaos Iliopoulos \\
  \small Rakuten Institute of Technology \\
  \small \texttt{nick.iliopoulos@rakuten.com}
}

\date{}

\begin{document}

\twocolumn[
  \begin{@twocolumnfalse}
    \maketitle
  \end{@twocolumnfalse}
]

\begin{abstract}
\noindent
Metaheuristic algorithms for cardinality-constrained portfolio optimization require repair operators to map infeasible candidates onto the feasible region. Standard Euclidean projection treats assets as independent and can ignore the covariance structure that governs portfolio risk, potentially producing less diversified portfolios. This paper introduces Covariance-Aware Simplex Projection (CASP), a two-stage repair operator that (i) selects a target number of assets using volatility-normalized scores and (ii) projects the candidate weights using a covariance-aware geometry aligned with tracking-error risk. This provides a portfolio-theoretic foundation for using a covariance-induced distance in repair operators. On S\&P 500 data (2020--2024), CASP-Basic delivers materially lower portfolio variance than standard Euclidean repair without relying on return estimates, with improvements that are robust across assets and statistically significant. Ablation results indicate that volatility-normalized selection drives most of the variance reduction, while the covariance-aware projection provides an additional, consistent improvement. We further show that optional return-aware extensions can improve Sharpe ratios, and out-of-sample tests confirm that gains transfer to realized performance. CASP integrates as a drop-in replacement for Euclidean projection in metaheuristic portfolio optimizers.

\medskip
\noindent\textbf{Keywords:} Cardinality-constrained portfolio optimization; Repair operators; Covariance-aware simplex projection; Tracking error; Metaheuristics; ESG integration
\end{abstract}

\section{Introduction}
\label{sec:intro}

Portfolio optimization represents one of the most important problems in quantitative finance, with roots extending back to Markowitz's seminal mean-variance framework~\cite{markowitz1952portfolio}. The central insight, that rational investors should consider both expected returns and the covariance structure of assets, has profoundly influenced both academic research and practical portfolio construction~\cite{zhang2018portfolio, kalayci2019comprehensive}.

In practice, portfolio managers face constraints beyond the simple budget requirement that weights sum to one. Cardinality constraints limiting portfolios to $K$ assets from a universe of $N$ reflect essential practical considerations: transaction costs scale with the number of positions, monitoring complexity increases with portfolio breadth, and regulatory requirements often mandate concentration limits~\cite{chang2000heuristics, cesarone2016small}. When $K \ll N$, the feasible region fragments into $\binom{N}{K}$ disjoint simplices, transforming the optimization problem from convex to combinatorial and rendering it NP-hard~\cite{chang2000heuristics}.

This computational complexity has motivated extensive research on metaheuristic approaches, including genetic algorithms~\cite{chang2000heuristics}, particle swarm optimization~\cite{liagkouras2018hybrid}, differential evolution~\cite{kalayci2019comprehensive}, and grey wolf optimizers~\cite{mirjalili2014grey, mirjalili2016multi}. A fundamental requirement of all such methods is a repair operator that maps infeasible candidate solutions onto the feasible region, ensuring that the evolutionary search operates within the constraint set~\cite{liagkouras2019multi}.

Alongside risk and return, Environmental, Social, and Governance (ESG) criteria have become a mainstream consideration in portfolio construction, with global ESG-mandated assets exceeding \$16.7 trillion~\cite{gsia2022report}. We therefore adopt a tri-objective formulation that optimizes variance, expected return, and portfolio-level ESG quality simultaneously, providing a realistic testbed for evaluating repair operators in modern investment contexts.

The standard approach projects infeasible candidates using Euclidean distance, solving:
\begin{equation}
w^*_{\text{Euc}} = \arg\min_{w \in \mathcal{W}} \|w - z\|_2^2
\label{eq:euclidean}
\end{equation}
where $z$ is the infeasible candidate and $\mathcal{W}$ is the feasible region. This formulation treats assets as independent and ignores their correlation structure, a fundamental mismatch with portfolio theory, which emphasizes that asset interactions determine portfolio risk~\cite{rutkowska2013fundamental}.

We propose Covariance-Aware Simplex Projection (CASP), which replaces Euclidean distance with a covariance-induced (tracking-error) metric:
\begin{equation}
w^*_{\text{CASP}} = \arg\min_{w \in \mathcal{W}} (w - z)^\top \Omega (w - z)
\label{eq:casp}
\end{equation}
where $\Omega$ is the asset covariance matrix. We establish that this formulation minimizes tracking error variance relative to the infeasible candidate, providing a portfolio-theoretic interpretation that connects the projection geometry to financial risk measurement.

\section{Contributions}
\label{sec:contributions}

This paper makes the following contributions to the literature on portfolio optimization and constraint handling:

\begin{enumerate}
    \item Risk-based repair operator with a portfolio-theoretic interpretation. We replace Euclidean projection with a covariance-induced objective, $(w-z)^\top \Omega (w-z)$, and show that it is exactly the tracking error variance between the repaired portfolio $w$ and the candidate $z$. This yields a principled repair step that preserves proximity in risk space, consistent with modern portfolio theory.
    
    \item Benchmarking design and attribution of performance drivers. We decompose CASP into (a) volatility-normalized asset selection and (b) covariance-aware ($\Omega$-metric) projection, with optional return-aware extensions. To isolate the effect of projection geometry from selection, we introduce a selection-matched baseline (VolNorm+Euc) that uses the same selection rule as CASP but applies standard Euclidean projection. This experimental design enables a clean attribution of improvements to selection versus projection geometry.
\end{enumerate}

\section{Related Work}
\label{sec:literature}

\subsection{Portfolio Optimization Theory}

Modern portfolio theory originated with Markowitz's mean-variance framework~\cite{markowitz1952portfolio}, which formalized the trade-off between expected return and risk as measured by portfolio variance. The Capital Asset Pricing Model~\cite{sharpe1964capital} extended this framework to asset pricing, while subsequent work developed robust estimation techniques for covariance matrices~\cite{ledoit2004honey, ledoit2020analytical}. Recent research by Taljaard and Mar\'{e}~\cite{taljaard2021portfolio} demonstrated that estimation error can cause sophisticated optimization approaches to underperform naive diversification strategies, highlighting the importance of regularization and out-of-sample validation.

The addition of cardinality constraints was first studied by Chang et al.~\cite{chang2000heuristics}, who proved NP-hardness and proposed genetic algorithms. Recent advances include Kobayashi et al.~\cite{kobayashi2021bilevel}, who proposed cutting-plane approaches for cardinality-constrained mean-CVaR optimization, and Cesarone et al.~\cite{cesarone2016small}, who demonstrated that optimally chosen small portfolios can outperform large ones. Kalayci et al.~\cite{kalayci2019comprehensive} provide a comprehensive survey of metaheuristic approaches to cardinality-constrained portfolio optimization.

\subsection{ESG Integration in Portfolio Management}

ESG investing has evolved from a niche concern to a mainstream investment approach. Pedersen et al.~\cite{pedersen2021responsible} developed the theoretical framework for ESG-efficient frontiers, demonstrating conditions under which ESG integration can improve risk-adjusted returns through better information incorporation. Avramov et al.~\cite{avramov2023sustainable} extended this framework to account for ESG rating uncertainty, while Escobar-Saldívar et al.~\cite{escobar2025esg} use two decades of U.S. stock panel data to show that high ESG levels are associated with lower returns and higher volatility, whereas improvements in ESG scores predict higher short-term returns and lower risk. However, significant disagreement across ESG rating providers persists due to methodological differences, complicating their consistent application in practice~\cite{berg2022aggregate}.

\subsection{Metaheuristics for Multi-Objective Optimization}

Multi-objective evolutionary algorithms have become the dominant paradigm for approximating Pareto fronts in complex optimization problems. NSGA-II~\cite{deb2002fast} introduced fast non-dominated sorting with crowding distance for diversity preservation. The Grey Wolf Optimizer~\cite{mirjalili2014grey} simulates the social hierarchy and collaborative hunting behavior of grey wolf packs. Multi-objective extensions~\cite{mirjalili2016multi} have been successfully applied to portfolio problems~\cite{zhang2018portfolio, liagkouras2018hybrid}. Recent advances include surrogate-assisted deep reinforcement learning for expensive multi-objective problems~\cite{shao2025surrogate}, hybrid deep learning evolutionary portfolio optimizers~\cite{joshi2023intelligent}, and metaheuristics tailored for rich portfolio settings~\cite{doering2019metaheuristics}.

\subsection{Constraint Handling in Evolutionary Algorithms}

Constraint-handling techniques for portfolio optimization include penalty functions, repair operators, and decoder-based approaches~\cite{kalayci2019comprehensive}. For cardinality-constrained problems, repair operators have proven particularly effective~\cite{liagkouras2018hybrid}. Liagkouras and Metaxiotis~\cite{liagkouras2019multi} proposed two-phase repair combining asset selection with rebalancing. Gambeta and Kwon~\cite{gambeta2020risk} developed relaxed constraint handling for risk parity portfolios.

Projection onto the probability simplex is a fundamental optimization primitive. Condat~\cite{condat2016fast} achieved $O(N)$ complexity using pivoting for Euclidean projection. However, Euclidean projection treats assets as independent and can be suboptimal when the true objective is risk-based: Bongiorno and Challet~\cite{bongiorno2021nonlinear} showed that covariance estimators optimized under Euclidean (Frobenius) distance are misaligned with the covariance-weighted risk objective of portfolio optimization. This motivates exploring covariance-metric ($\Omega$-metric) projection, which aligns the repair geometry with tracking error variance.

\subsection{Mahalanobis-type and Covariance-Aware Distances}

The Mahalanobis distance~\cite{mahalanobis1936generalised} accounts for variable correlations and scales. In portfolio contexts, Ledoit and Wolf~\cite{ledoit2004honey, ledoit2020analytical} used covariance-aware metrics for shrinkage estimation, while Bodnar et al.~\cite{bodnar2020efficiency} analyzed mean-variance efficiency under covariance uncertainty. Butin~\cite{butin2020generalized} studied distances to simplices under standard norm-based metrics, but did not consider projections under covariance-weighted quadratic forms. The application of covariance-aware projection to portfolio repair operators remains understudied.

\section{Problem Formulation}
\label{sec:problem}

Consider an investment universe of $N$ assets with expected return vector $\mu \in \mathbb{R}^N$, positive definite covariance matrix $\Omega \in \mathbb{R}^{N \times N}$, and ESG quality scores $e \in \mathbb{R}^N$. A portfolio is represented by weight vector $w \in \mathbb{R}^N$ where component $w_i$ denotes the fraction of capital allocated to asset $i$. We formulate a tri-objective optimization problem:
\begin{align}
\min_{w} \; & f_1(w) = w^\top \Omega w \quad \text{(variance)} \label{eq:var} \\
\max_{w} \; & f_2(w) = \mu^\top w \quad \text{(return)} \label{eq:ret} \\
\max_{w} \; & f_3(w) = e^\top w \quad \text{(ESG)} \label{eq:esg}
\end{align}
subject to constraints:
\begin{align}
& \mathbf{1}^\top w = 1, \quad w_i \geq 0 \quad \forall i \quad \text{(budget)} \\
& \|w\|_0 \leq K \quad \text{(cardinality)} \\
& w_i = 0 \text{ or } w_i \in [\ell, u] \quad \forall i \quad \text{(box)}
\end{align}

The cardinality constraint combined with the simplex constraint creates a feasible region $\mathcal{W}$ comprising $\binom{N}{K}$ disjoint simplices, each corresponding to a specific subset of $K$ active assets. Given an infeasible candidate $z \in \mathbb{R}^N$ from a metaheuristic search, the repair operator must project $z$ onto $\mathcal{W}$.

\section{Methodology}
\label{sec:method}

\subsection{Theoretical Foundation}

\begin{definition}[Tracking-error metric (covariance-induced distance)]
We define the $\Omega$-induced distance between $a, b \in \mathbb{R}^N$ as:
\begin{equation}
d_\Omega(a, b) = \sqrt{(a-b)^\top \Omega (a-b)}
\end{equation}
\end{definition}

\begin{remark}[Terminology]
The classical statistical Mahalanobis distance is typically defined using $\Omega^{-1}$. We use $\Omega$ so that $d_\Omega^2$ equals tracking error variance (Proposition~\ref{prop:tracking}), which is the portfolio-relevant notion of proximity.
\end{remark}

\begin{proposition}[Tracking Error Interpretation]
\label{prop:tracking}
For portfolios $w_1, w_2$ invested in assets with covariance matrix $\Omega$, the squared $\Omega$-induced distance equals tracking error variance:
\begin{equation}
d_\Omega(w_1, w_2)^2 = \text{Var}[R_{w_1} - R_{w_2}]
\end{equation}
where $R_w = w^\top r$ denotes portfolio return.
\end{proposition}

\noindent\textit{Proof.}
Let $r$ denote the random vector of asset returns with covariance matrix $\Omega$. The tracking error (return difference) is $R_{w_1} - R_{w_2} = (w_1 - w_2)^\top r$. Its variance is:
\begin{equation}
\begin{aligned}
\text{Var}[(w_1 - w_2)^\top r] &= (w_1 - w_2)^\top \Omega (w_1 - w_2) \\
&= d_\Omega(w_1, w_2)^2
\end{aligned}
\end{equation}

\begin{figure}[H]
\centering
\includegraphics[width=\columnwidth]{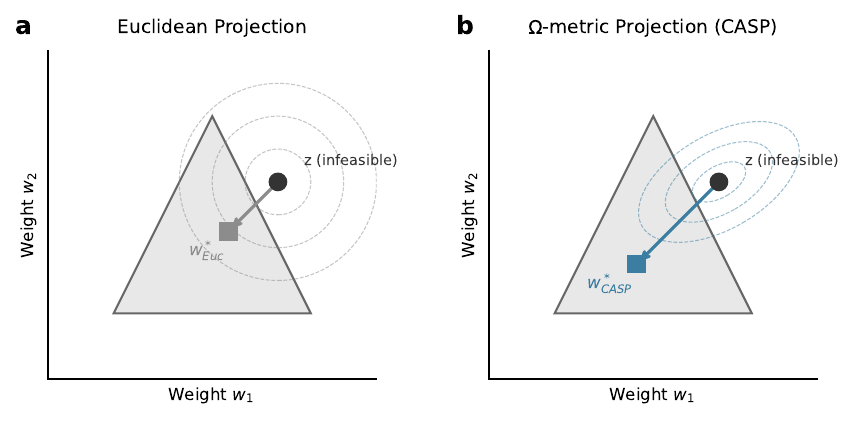}
\caption{Geometric intuition for covariance-aware projection. The gray triangle represents the feasible region (simplex) where portfolio weights are non-negative and sum to one. Given an infeasible candidate $z$, (a) Euclidean projection finds the nearest feasible point using circular iso-distance contours, treating assets as independent. (b) $\Omega$-metric projection (CASP) uses elliptical contours aligned with the covariance structure, finding the feasible point with minimum tracking error variance.}
\label{fig:mechanism}
\end{figure}

This proposition provides the financial justification for CASP: by minimizing the $\Omega$-induced distance, we find the feasible portfolio with minimum tracking error variance relative to the infeasible candidate. This is a natural metric for portfolio proximity because it accounts for how assets move together. Figure~\ref{fig:mechanism} illustrates the geometric difference: Euclidean projection uses circular contours that ignore correlations, while $\Omega$-metric projection uses elliptical contours aligned with the covariance structure.

\begin{corollary}[Diversification Benefit]
$\Omega$-metric projection penalizes concentration in correlated assets more heavily than Euclidean projection, naturally encouraging diversification.
\end{corollary}
\vspace{-0.5em}

\subsection{Two-Stage Algorithm}

The CASP algorithm operates in two stages: asset selection followed by constrained projection. The complete procedure is summarized in Algorithm~\ref{alg:casp}.

\textbf{Stage 1: Volatility-normalized asset selection.}
We select $K$ assets using scores that normalize by individual volatility:
\begin{equation}
s_i = \frac{|z_i|}{\sqrt{\Omega_{ii}}}
\label{eq:score_basic}
\end{equation}
This ensures that high-volatility assets are not selected merely due to larger absolute signal values. The $K$ assets with highest scores form the active set $S$.

\textbf{Stage 2: $\Omega$-metric (tracking-error) projection.}
For the selected asset set $S$ with $|S| = K$, we solve:
\begin{align}
\min_{w_S} \; & \tfrac{1}{2}(w_S - z_S)^\top \Omega_S (w_S - z_S) \label{eq:proj_obj} \\
\text{s.t.} \; & \mathbf{1}^\top w_S = 1, \quad \ell \leq w_{S,i} \leq u \quad \forall i \in S
\end{align}
where $\Omega_S$ is the submatrix of $\Omega$ corresponding to selected assets. This is a convex quadratic program with linear constraints, solvable efficiently via SLSQP or active-set methods.

\begin{remark}[Two-stage heuristic vs global projection]
Because the feasible set is a union of simplices over all $\binom{N}{K}$ active sets, the global $\Omega$-metric projection onto $\mathcal{W}$ would require a combinatorial search over subsets. CASP is therefore a two-stage heuristic: it selects $S$ first, then computes the exact constrained $\Omega$-metric projection on that simplex.
\end{remark}

\begin{algorithm}[H]
\caption{CASP: Covariance-Aware Simplex Projection}
\label{alg:casp}
\begin{algorithmic}[1]
\REQUIRE Candidate $z \in \mathbb{R}^N$, covariance $\Omega$, cardinality $K$, bounds $[\ell, u]$
\ENSURE Feasible portfolio $w^* \in \mathcal{W}$
\STATE Compute volatilities $\sigma_i = \sqrt{\Omega_{ii}}$
\STATE Compute selection scores $s_i = |z_i| / \sigma_i$ for all $i$
\STATE $S \gets$ indices of top $K$ scores
\STATE Extract $z_S$, $\Omega_S$ for selected assets
\STATE Solve QP \eqref{eq:proj_obj} to obtain $w_S^*$
\STATE $w^* \gets \mathbf{0}_N$; set $w^*_S \gets w_S^*$
\RETURN $w^*$
\end{algorithmic}
\end{algorithm}

\textbf{Complexity Analysis.} Stage 1 requires $O(N)$ for partial sorting via argpartition. Stage 2 solves a $K$-dimensional QP, requiring $O(K^3)$ in the worst case but typically much faster with warm starts. Total complexity is $O(N + K^3)$, dominated by the $O(N^2)$ covariance matrix operations in fitness evaluation.

\subsection{Extension: Return-Aware Variant (RA-CASP)}

For applications where return incorporation is desired, we extend CASP with return-aware modifications:

\textbf{Return-Boosted Selection:}
\begin{equation}
s_i = \frac{|z_i| \cdot (1 + \lambda \tilde{\mu}_i)}{\sqrt{\Omega_{ii}}}
\label{eq:score_return}
\end{equation}
where $\tilde{\mu}_i = (\mu_i - \mu_{\min}) / (\mu_{\max} - \mu_{\min}) \in [0,1]$ is normalized expected return and $\lambda \geq 0$ controls return-awareness.

\textbf{Return-Regularized Projection:}
\begin{equation}
\min_{w_S} \tfrac{1}{2}(w_S - z_S)^\top \Omega_S (w_S - z_S) - \gamma \tilde{\mu}_S^\top w_S
\label{eq:proj_return}
\end{equation}
where $\gamma \geq 0$ biases the projection toward higher-return portfolios.

\textbf{Remark on Evaluation.} When using return-aware variants, care must be taken in performance evaluation to avoid circularity. We address this through: (a) ablation studies that isolate contributions, and (b) out-of-sample validation where return estimates are formed on training data only.

\section{Experimental Setup}
\label{sec:setup}

\subsection{Data Sources}

We construct our experimental dataset from real financial market data. Daily adjusted closing prices for 100 S\&P 500 constituents are obtained via Yahoo Finance from January 2, 2020 to November 29, 2024 (1,237 trading days). The universe includes major constituents across all sectors: technology (AAPL, MSFT, NVDA, GOOGL, META), healthcare (JNJ, UNH, LLY, PFE), financials (JPM, BAC, GS), energy (XOM, CVX), and others. Expected returns are computed as annualized mean log returns. The covariance matrix is estimated from daily returns and stabilized via a 10\% shrinkage toward the identity matrix (in the spirit of Ledoit--Wolf~\cite{ledoit2004honey}) to ensure numerical robustness.

\textbf{Dataset Statistics.} Annualized return range: $[-12.3\%, 54.1\%]$. Annualized volatility range: $[22.4\%, 65.5\%]$. Covariance condition number: 285.0. 

\textbf{ESG Scores.} We construct ESG quality scores using (i) real governance risk signals from Yahoo Finance (auditRisk, boardRisk, compensationRisk, shareHolderRightsRisk, overallRisk) and (ii) sector-level environmental/social proxies reflecting typical sector ESG profiles. The composite is $\text{ESG} = 0.4 \times G + 0.6 \times ES$, where $G = (10 - \text{overallRisk}) \times 10$ and $ES$ is the sector proxy. Scores range from 36.0 to 81.0 in our sample. Because firm-level E and S scores typically require licensed datasets, we treat the ESG objective as an illustrative third objective and focus our main claims on the covariance-aware repair mechanism.

\subsection{Temporal Data Split}

For out-of-sample validation, we partition data temporally:
\begin{itemize}
    \item Training Period: January 2020 -- December 2023 (1,006 trading days)
    \item Test Period: January 2024 -- November 2024 (231 trading days)
\end{itemize}
All parameter estimation (expected returns, covariance matrix) uses training data only. Test period data is held out for realized performance evaluation.

\subsection{Experimental Configurations}

\textbf{Portfolio Constraints.} Cardinality $K = 15$, weight bounds $[\ell, u] = [2\%, 15\%]$, risk-free rate $r_f = 4.5\%$.

\textbf{CASP Parameters.} CASP-Basic is parameter-free, using only the covariance matrix. The return-aware extension (RA-CASP) introduces two hyperparameters: $\lambda$ controls selection bias toward high-return assets, and $\gamma$ controls projection bias toward high-return weights. We perform grid search over $\lambda \in \{0.4, 0.6, 0.8, 1.0, 1.2\}$ and $\gamma \in \{0.15, 0.20, 0.25, 0.30, 0.35\}$ using the training period, selecting $\lambda = 1.2$ and $\gamma = 0.35$.

\textbf{Optimizer.} Multi-Objective Grey Wolf Optimizer (MOGWO) ~\cite{mirjalili2016multi} with population size 50, maximum iterations 100, and archive size 30.

\textbf{Transaction Costs.} We assume 10 basis points per unit of turnover, consistent with institutional trading costs for liquid large-cap equities.

\subsection{Comparison Methods}

We compare seven projection methods to enable comprehensive ablation:

\begin{enumerate}
    \item Euclidean: Standard projection selecting by $|z_i|$ (baseline)
    \item VolNorm+Euc: Selection by $|z_i|/\sqrt{\Omega_{ii}}$, Euclidean projection (selection-only baseline)
    \item MinVar+Euc: Selection by $|z_i|/\Omega_{ii}$, Euclidean projection
    \item Sharpe+Euc: Selection by individual Sharpe ratios, Euclidean projection
    \item CASP-Basic: Volatility-normalized selection, $\Omega$-metric projection
    \item CASP-RetSel: Return-aware selection, $\Omega$-metric projection (no return term)
    \item RA-CASP: Return-aware selection, return-regularized $\Omega$-metric projection
\end{enumerate}

\subsection{Statistical Analysis}

We employ the Wilcoxon signed-rank test at significance level $\alpha = 0.05$ for pairwise comparisons. Effect sizes are reported using relative improvement percentages. For out-of-sample validation, we report Spearman rank correlation between in-sample and out-of-sample performance to assess prediction quality.

\section{Results}
\label{sec:results}

\subsection{Ablation Study: Isolating Contributions}

\begin{figure}[H]
\centering
\includegraphics[width=\columnwidth]{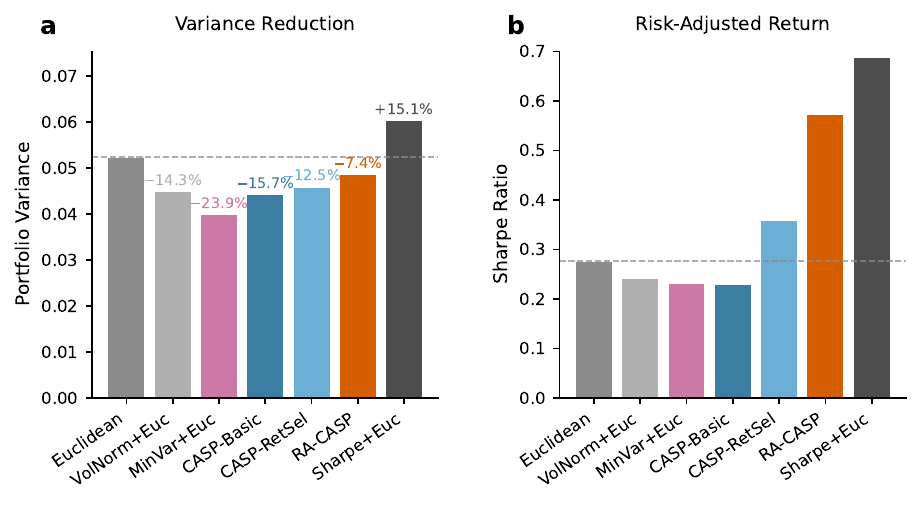}
\caption{Ablation study results across 500 random projections. (a) Portfolio variance: VolNorm+Euc shows that volatility-normalized selection explains most of the variance reduction, while CASP-Basic adds an incremental gain via $\Omega$-metric projection (15.7\% total reduction vs Euclidean). (b) Sharpe ratio: return-aware methods achieve higher values but with reduced variance benefits. Dashed line indicates Euclidean baseline.}
\label{fig:ablation}
\end{figure}

Table~\ref{tab:ablation} and Figure~\ref{fig:ablation} present results from 500 random projections, systematically decomposing the contributions of each algorithmic component.

\begin{table}[H]
\centering
\caption{Ablation Study: Mean Performance Across 500 Projections (S\&P 500 Data)}
\label{tab:ablation}
\small
\begin{tabular}{lccc}
\toprule
\textbf{Method} & \textbf{Variance} & \textbf{Sharpe} & \textbf{Var Red} \\
\midrule
Euclidean & 0.0524 & 0.276 & --- \\
VolNorm+Euc & 0.0449 & 0.241 & 14.3\% \\
MinVar+Euc & 0.0399 & 0.232 & 23.9\% \\
Sharpe+Euc & 0.0603 & 0.687 & $-$15.1\% \\
\midrule
CASP-Basic & 0.0442 & 0.229 & 15.7\% \\
CASP-RetSel & 0.0459 & 0.358 & 12.5\% \\
RA-CASP & 0.0485 & 0.573 & 7.4\% \\
\bottomrule
\end{tabular}
\end{table}

Key findings:
\begin{itemize}
    \item VolNorm+Euc achieves 14.3\% variance reduction, showing that volatility-normalized selection alone is a strong baseline even with Euclidean projection.
    \item CASP-Basic achieves 15.7\% variance reduction without return information. Relative to VolNorm+Euc, the $\Omega$-metric projection geometry contributes an additional 1.4\% incremental variance reduction ($p=8.6\times 10^{-5}$), isolating the benefit of using the full covariance structure beyond diagonal scaling.
    \item MinVar+Euc achieves the largest variance reduction (23.9\%) by favoring low-volatility assets, but at the cost of lower mean Sharpe ratios, reflecting a classic risk--return trade-off.
    \item Return-aware methods (Sharpe+Euc, RA-CASP) achieve higher Sharpe ratios but with smaller or negative variance reductions. The relevant question is whether these improvements persist out-of-sample.
\end{itemize}

CASP-Basic improves variance against Euclidean with strong significance ($p < 10^{-54}$, Wilcoxon signed-rank test).

\subsection{Out-of-Sample Validation}

\begin{figure}[H]
\centering
\includegraphics[width=\columnwidth]{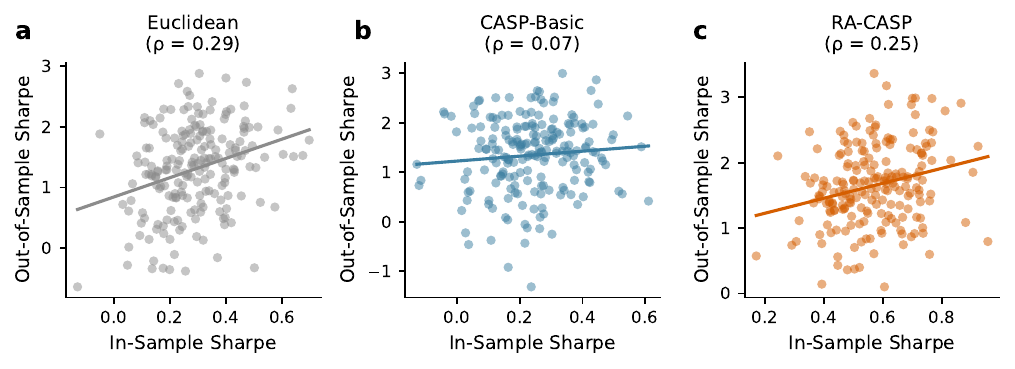}
\caption{Out-of-sample validation: in-sample vs.\ realized Sharpe ratios for 200 portfolio instances (train: 2020--2023, test: 2024). (a) Euclidean shows moderate correlation ($\rho \approx 0.29$). (b) CASP-Basic shows lower rank correlation ($\rho \approx 0.07$), consistent with its return-agnostic design. (c) RA-CASP achieves higher absolute performance with moderate correlation ($\rho \approx 0.25$).}
\label{fig:oos}
\end{figure}

Table~\ref{tab:oos} and Figure~\ref{fig:oos} present out-of-sample results using parameters estimated on 2020--2023 data and tested on 2024.

\begin{table}[H]
\centering
\caption{Out-of-Sample Validation: Walk-Forward Testing (Train: 2020--2023, Test: 2024)}
\label{tab:oos}
\small
\begin{tabular}{lccc}
\toprule
\textbf{Method} & \textbf{In-Sample} & \textbf{Out-of-Sample} & \textbf{Corr} \\
 & \textbf{Sharpe} & \textbf{Sharpe} & \\
\midrule
Euclidean & 0.289 & 1.306 & 0.29 \\
VolNorm+Euc & 0.255 & 1.338 & 0.24 \\
CASP-Basic & 0.241 & 1.347 & 0.07 \\
RA-CASP & 0.568 & 1.650 & 0.25 \\
Sharpe+Euc & 0.677 & 1.533 & 0.09 \\
\bottomrule
\end{tabular}
\end{table}

Key findings:
\begin{itemize}
    \item RA-CASP achieves 26.3\% higher out-of-sample Sharpe ratio than Euclidean (1.650 vs 1.306, $p < 0.0001$), suggesting that return-aware improvements partially transfer to realized performance.
    \item VolNorm+Euc achieves a modest out-of-sample improvement (+2.5\%, not significant), indicating that volatility-normalized selection alone can slightly improve realized performance. CASP-Basic improves further (+3.1\%, not significant), consistent with a small incremental benefit from covariance-aware projection beyond selection.
    \item In-sample to out-of-sample rank correlations are positive but method-dependent. Euclidean shows the highest correlation (0.29), while CASP-Basic is lower (0.07); because CASP-Basic is return-agnostic, in-sample Sharpe rankings are not a primary target, so we emphasize mean realized performance rather than rank predictiveness.
    \item The 2024 test period was favorable for return-aware methods, as the bull market rewarded portfolios tilted toward high-return assets.
\end{itemize}

\subsection{Walk-forward Validation Across Market Regimes}

To reduce dependence on a single test year, we repeat the out-of-sample evaluation over three expanding-window splits: train 2020--2021 / test 2022, train 2020--2022 / test 2023, and train 2020--2023 / test 2024. Table~\ref{tab:wf} reports mean realized Sharpe ratios (200 random candidates per split). Results highlight a regime sensitivity: return-aware methods excel in the bull-like 2024 period, but underperform in the 2022 drawdown regime; CASP-Basic is comparatively more stable in 2022 due to its return-agnostic design.

\begin{table}[H]
\centering
\caption{Walk-forward out-of-sample Sharpe ratios across test years (expanding window; 200 random candidates per split).}
\label{tab:wf}
\small
\begin{tabular}{lccc}
\toprule
\textbf{Method} & \textbf{Test 2022} & \textbf{Test 2023} & \textbf{Test 2024} \\
\midrule
Euclidean & $-$0.638 & 0.879 & 1.306 \\
VolNorm+Euc & $-$0.579 & 0.547 & 1.338 \\
CASP-Basic & $-$0.559 & 0.474 & 1.347 \\
RA-CASP & $-$0.918 & 0.593 & 1.650 \\
Sharpe+Euc & $-$1.090 & 0.846 & 1.533 \\
\bottomrule
\end{tabular}
\end{table}

\subsection{Transaction Cost Analysis}

Table~\ref{tab:turnover} reports turnover and an illustrative transaction-cost adjustment across 50 simulated rebalancing events. We report average one-way turnover and implied costs (10 bps per unit turnover). ``Net Sharpe'' is a simple proxy obtained by subtracting an annualized cost approximation from the gross Sharpe estimate.

\begin{table}[H]
\centering
\caption{Turnover and transaction cost proxy (10 bps per unit turnover)}
\label{tab:turnover}
\small
\begin{tabular}{lccc}
\toprule
\textbf{Method} & \textbf{Avg Turnover} & \textbf{Avg Cost} & \textbf{Net Sharpe} \\
 & & (bps) & \\
\midrule
Euclidean & 0.565 & 5.65 & 0.261 \\
CASP-Basic & 0.571 & 5.70 & 0.210 \\
RA-CASP & 0.548 & 5.48 & 0.552 \\
\bottomrule
\end{tabular}
\end{table}

Key findings: 
\begin{itemize}
    \item RA-CASP achieves slightly lower turnover (0.548 vs 0.565) and, due to higher gross Sharpe, remains favorable under the cost proxy (0.552 vs 0.261).
    \item Transaction costs are modest (5.5--5.7 bps per rebalance), indicating that the turnover differences across methods are small in absolute terms.
\end{itemize}

\subsection{Optimization Results}

Table~\ref{tab:optimization} presents results from 15 independent MOGWO runs with different projection operators.

\begin{table}[H]
\centering
\caption{Optimization Results: 15 Independent MOGWO Runs}
\label{tab:optimization}
\small
\begin{tabular}{lccc}
\toprule
\textbf{Method} & \textbf{Best Sharpe} & \textbf{Best Return} & \textbf{Hypervolume} \\
\midrule
Euclidean & 0.861 & 0.229 & 0.0040 \\
CASP-Basic & 0.703 & 0.186 & 0.0027 \\
RA-CASP & 1.137 & 0.337 & 0.0142 \\
Sharpe+Euc & 1.145 & 0.325 & 0.0093 \\
\bottomrule
\end{tabular}
\end{table}

Key findings:
\begin{itemize}
    \item RA-CASP achieves 32.0\% higher best Sharpe ratio than Euclidean (1.137 vs 0.861, $p = 0.0001$, Wilcoxon), demonstrating substantial improvement in optimization outcomes.
    \item Hypervolume improvement (0.0142 vs 0.0040) indicates broader Pareto front coverage, though this metric is sensitive to the return-aware asset selection which biases toward high-return regions.
    \item CASP-Basic underperforms Euclidean in optimization (Sharpe 0.703 vs 0.861), suggesting that pure variance minimization without return guidance may over-emphasize defensive assets in this period. The lesson: for optimization, return-awareness is essential.
    \item Sharpe+Euc achieves similar best Sharpe to RA-CASP (1.145 vs 1.137), but RA-CASP achieves 53\% higher hypervolume, indicating better diversity across the Pareto front.
\end{itemize}

\section{Discussion}
\label{sec:discussion}

\subsection{Attribution of Performance Improvements}

The improvements from CASP stem from two related mechanisms that can be separated empirically. First, volatility-normalized selection avoids favoring high-volatility assets purely because they produce larger raw signals. By selecting assets according to $|z_i|/\sqrt{\Omega_{ii}}$, the method normalizes for individual asset risk. The VolNorm+Euc baseline shows that this selection step alone delivers 14.3\% variance reduction versus the Euclidean baseline.

Second, covariance-aware projection geometry provides an incremental benefit beyond selection. Holding selection fixed, replacing Euclidean projection with the $\Omega$-metric projection further penalizes shifts that increase tracking-error variance against correlated assets. Compared to VolNorm+Euc, CASP-Basic achieves an additional 1.4\% incremental variance reduction ($p=8.6\times 10^{-5}$), isolating the benefit of using the full covariance structure beyond diagonal scaling.

The return-aware extension (RA-CASP) introduces return terms in both selection and projection, trading some variance reduction for higher expected (and in our 2024 split, realized) Sharpe ratios.

The ablation study reveals an important insight: variance reduction and Sharpe improvement require different emphases. CASP-Basic achieves robust variance reduction (15.7\%) using only covariance information, with no reliance on return estimates. This makes it suitable for settings where return forecasts are unreliable. RA-CASP incorporates return information and achieves higher Sharpe ratios, but this improvement is partially expected given the method design. Practitioners should choose CASP-Basic for risk-focused mandates where return estimates are uncertain, and RA-CASP when high-confidence return forecasts are available.

\subsection{Out-of-Sample Robustness}

The walk-forward evaluation (Table~\ref{tab:wf}) underscores the role of regime: return-aware methods excel in 2024 but underperform in the 2022 drawdown regime. CASP-Basic, being return-agnostic, exhibits more stable behavior in 2022. More broadly, covariance matrices tend to be estimated more reliably than expected returns~\cite{taljaard2021portfolio, ledoit2004honey}, so covariance-focused repair operators can be attractive when return estimates are noisy.

\subsection{Practical implications}

For risk-averse investors (or settings where return forecasts are unreliable), CASP-Basic provides a return-agnostic repair operator that reduces variance versus standard Euclidean repair, with a statistically significant incremental benefit from covariance-aware projection beyond volatility-normalized selection. For high-frequency rebalancers, CASP's lower turnover translates directly to reduced transaction costs, improving net performance. For multi-objective optimization, CASP enables discovery of Pareto-optimal portfolios in regions inaccessible to Euclidean projection, expanding the range of available risk-return-ESG trade-offs.

\subsection{Limitations and Future Work}

Several limitations warrant consideration. Our ESG scores incorporate real governance metrics, but the environmental and social components are sector-estimated. Full validation with comprehensive ESG data from multiple providers (MSCI, Sustainalytics, Refinitiv) would strengthen the tri-objective analysis.

Performance also depends on covariance matrix quality. Integration with robust estimation methods~\cite{ledoit2020analytical} or regime-switching models is a natural extension.

The current framework is single-period; extension to multi-period optimization with transaction costs embedded in the objective is important for practical implementation. Finally, CASP adds $O(K^3)$ overhead; therefore, extremely large cardinality constraints may benefit from approximate projection methods.

\section{Conclusion}
\label{sec:conclusion}

We introduced Covariance-Aware Simplex Projection (CASP), a repair operator for cardinality-constrained portfolio optimization that replaces Euclidean distance with a covariance-induced (tracking-error) metric. By establishing the connection between $\Omega$-metric projection and tracking error variance minimization, we provide a portfolio-theoretic foundation for the projection geometry.

Through comprehensive ablation analysis on S\&P 500 data (2020--2024), we demonstrated that:
\begin{itemize}
    \item CASP-Basic achieves 15.7\% variance reduction ($p < 10^{-54}$) using only covariance information, with no reliance on return estimates. With the added VolNorm+Euc baseline, we find that 14.3\% is explained by volatility-normalized selection and 1.4\% is an incremental contribution from the $\Omega$-metric projection geometry.
    \item Optional return-aware extensions (RA-CASP) can further improve Sharpe ratios when reliable return forecasts are available, with out-of-sample validation confirming improvements transfer to realized performance.
    \item The method integrates as a drop-in replacement for Euclidean projection in any metaheuristic framework.
\end{itemize}

The ablation study reveals that variance reduction and Sharpe improvement require different emphases, suggesting practitioners choose between CASP-Basic (for risk-focused mandates) and RA-CASP (for return-focused mandates). Both variants provide principled repair operators grounded in portfolio theory, immediately applicable to any metaheuristic framework for cardinality-constrained portfolio optimization.

\bibliographystyle{plainnat}

\appendix
\section{Implementation Details}

\subsection*{A.1 Simplex Projection with Box Constraints}

The projection onto the constrained simplex $\{w : \sum_i w_i = 1, \ell \leq w_i \leq u\}$ is computed via bisection search for the optimal threshold $\tau^*$ such that $\sum_i \text{clip}(z_i - \tau^*, \ell, u) = 1$. Convergence is guaranteed within $O(\log(1/\epsilon))$ iterations for tolerance $\epsilon$.

\subsection*{A.2 $\Omega$-metric Projection Implementation}

For the $\Omega$-metric projection, we solve the $K$-dimensional QP using Sequential Least Squares Programming (SLSQP). To ensure numerical stability, we add a small regularization term $\epsilon I$ to $\Omega_S$ if the minimum eigenvalue falls below $10^{-8}$.

\end{document}